\documentclass[11pt,twoside]{article}

\usepackage{asp2006}
\usepackage{epsf}
\usepackage{psfig}
\usepackage{lscape}

\begin{document}
\title{Extended UV (XUV) Emission in Nearby Galaxy Disks}   
\author{
Armando Gil de Paz\altaffilmark{1},
David A. Thilker\altaffilmark{2},
Luciana Bianchi\altaffilmark{2},
Alfonso Arag\'on-Salamanca\altaffilmark{3},
Samuel Boissier\altaffilmark{4},
Barry F. Madore\altaffilmark{5},
Cristina D\'{\i}az-L\'opez\altaffilmark{1},
Ignacio Trujillo\altaffilmark{6},
Michael Pohlen\altaffilmark{7},
Peter Erwin\altaffilmark{8},
Jaime Zamorano\altaffilmark{1},
Jes\'us Gallego\altaffilmark{1},
Jorge Iglesias-P\'aramo\altaffilmark{9},
Jos\'e M. V\'{\i}lchez\altaffilmark{9},
Mercedes Moll\'a\altaffilmark{10},
Juan Carlos Mu\~noz-Mateos\altaffilmark{1},
Pablo G. P\'erez-Gonz\'alez\altaffilmark{1},
Santos Pedraz\altaffilmark{11},
Kartik Sheth\altaffilmark{12},
Robert C. Kennicutt Jr.\altaffilmark{13},
Robert Swaters\altaffilmark{14},
and the GALEX Science Team
}   
\altaffiltext{1}{Departamento de Astrof\'{\i}sica, Universidad Complutense de Madrid, Madrid 28040, Spain}
\altaffiltext{2}{Center for Astrophysical Sciences, Johns Hopkins University, Baltimore, MD 21218}
\altaffiltext{3}{School of Physics and Astronomy, University of Nottingham, NG7 2RD, UK}
\altaffiltext{4}{Laboratoire d'Astrophysique de Marseille, 13376 Marseille Cedex 12, France}
\altaffiltext{5}{The Observatories, Carnegie Institution of Washington, Pasadena, CA 91101}
\altaffiltext{6}{Instituto de Astrof\'{\i}sica de Canarias, 38200 La Laguna, Tenerife, Spain}
\altaffiltext{7}{Cardiff University, School of Physics \& Astronomy,
  Cardiff, CF24 3AA, UK}
\altaffiltext{8}{Max-Planck-Institut f\"ur Extraterrestrische Physik, D-85748 Garching, Germany}
\altaffiltext{9}{Instituto de Astrof\'{\i}sica de Andaluc\'{\i}a (CSIC), C$.$ Bajo de Hu\'etor 50, 18008 Granada, Spain}
\altaffiltext{10}{Departamento de Investigaci\'on B\'asica, CIEMAT, 28040 Madrid, Spain}
\altaffiltext{11}{Centro Astron\'omico Hispano Alem\'an, Calar Alto (CSIC-MPI), 04004 Almer\'{\i}a, Spain}
\altaffiltext{12}{Spitzer Science Center, California Institute of Technology, Pasadena, CA 91125}
\altaffiltext{13}{Institute of Astronomy, University of Cambridge, Cambridge CB3 0HA, UK}
\altaffiltext{14}{Department of Astronomy, University of Maryland, College Park, MD 20742-2421}

\begin{abstract} 
We summarize the main properties of the extended UV (XUV) emission
found in roughly 30\% of the nearby spiral galaxies observed by
the GALEX satellite. Two different classes of XUV disks are
identified, the Type~1 XUV disks where significant, structured
UV-bright features are found beyond the {\it classical}
azimuthally-averaged star-formation threshold, and the Type~2 XUV
disks, which are characterized by very extended (seven times the area
where most of the stellar mass is found), blue [(FUV$-$$K$)$<$5\,mag]
outer disks. These latter disks are extreme examples of galaxies
growing inside-out. The few XUV disks studied in detail to date are
rich in HI but relatively poor in molecular gas, have stellar
populations with luminosity-weighted ages of $\sim$1\,Gyr, and
ionized-gas metal abundances of $\sim$Z$_{\odot}$/10. As part of
the CAHA-XUV project we are in the process of obtaining deep
multi-wavelength imaging and spectroscopy of 65 XUV-disk galaxies so
to determine whether or not these properties are common among XUV
disks.
\end{abstract}

\section*{Introduction}    
Observations by the GALEX satellite (Martin et al$.$ 2005) have
recently revealed the presence of UV-bright emission at large
galactocentric distances (up to four times the optical radius) in a
number of nearby spiral galaxies (see Figure~1). This phenomenon was
first seen in the outer regions of the disks of M~83 (Thilker et al$.$
2005) and NGC~4625 (Gil de Paz et al$.$ 2005). Previous evidence of
in-situ star formation in the outer disks of galaxies include Ferguson
et al$.$ (1998) and van Zee et al$.$ (1998). The existence of UV
emission at large radii is of great relevance since it questions the
mere existence of an azimuthally-averaged threshold for the star
formation and provides an important test for the applicability of the
Kennicutt-Schmidt law at low gas densities. The XUV phenomenon is also
useful as a test for the models of the formation and evolution of
disks in galaxies, both for SPH/N-body simulations (e.g$.$ Brook et
al$.$ 2006) and chemical and spectro-photometric "backward" models
(Boissier \& Prantzos 2000; Moll\'a et al$.$ 1996), and, more
specifically, for its predicted inside-out formation scenario
(Mu\~noz-Mateos et al$.$ 2007).

\begin{figure}[ht!]
\plotone{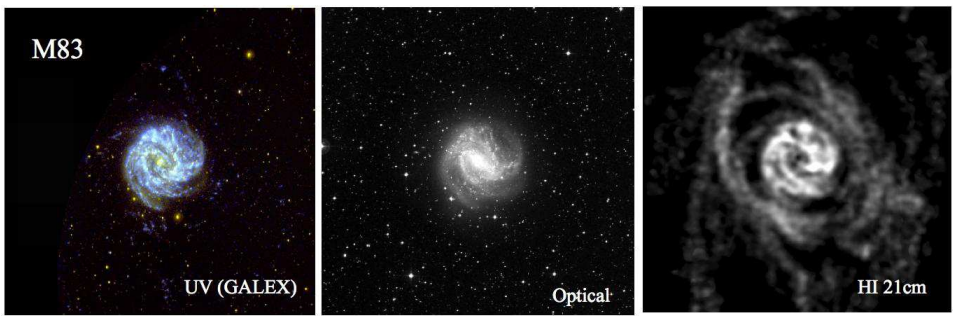}
\caption{Ultraviolet GALEX (RGB composite of the FUV and NUV images; {\it left}), 
optical Digitized Sky Survey ({\it center}), and HI 21\,cm ({\it right}) images of the XUV-disk galaxy M~83.}
\end{figure}

\section*{XUV-disks Classification}    
A study of the spiral galaxy population included in the {\it GALEX
  Atlas of Nearby Galaxies} (Gil de Paz et al$.$ 2007a) revealed that
$\sim$30\% of the disk galaxies in the local (D$<$40\,Mpc) Universe
show some degree of XUV emission (Thilker et al$.$ 2007). Roughly two
thirds of them (20\% of the total) show UV-bright complexes structured
in the form of spiral segments or irregularly-grouped features beyond
the {\it classical} threshold for star formation deduced from the
analysis of H$\alpha$ imaging data (Martin \& Kennicutt 2001), which
is located at
$\mu_{\mathrm{FUV}}$$\simeq$27.25\,AB\,mag\,arcsec$^{-2}$. These
galaxies are classified as Type 1 XUV disks (Thilker et al$.$ 2007).
In the remaining third the XUV emission is characterized by ubiquitous
UV emission in regions of faint or undetectable optical and
near-infrared emission [(FUV$-$$K$)$<$5\,mag] but inside the
surface-brightness threshold for star formation. In these Type 2 XUV
disks the blue outer disks cover an area which is at least seven times
larger than the area of the isophote encompassing 80\% of the light in
the $K$ band (i.e$.$ $\sim$80\% of the stellar mass). Type 2 objects
are examples of disks growing inside-out at a very high rate (see
Mu\~noz-Mateos et al$.$ 2007).

\section*{Photometric and Chemical Properties}    
The XUV disks are characterized by very blue UV-optical color
profiles. The comparison of these colors with the predictions of
evolutionary synthesis models indicates luminosity-weighted ages that
are either $\sim$1\,Gyr for a continuous star formation or younger if
the formation was instantaneous (Gil de Paz et al$.$ 2005, 2007b). We
cannot rule out the presence of a more evolved stellar population on
top of which significant star-formation activity is now taking place
with, perhaps, additional episodes of star formation in between. The
analysis of chemical abundances of the outermost regions in the XUV
disks of M~83 and NGC~4625 (Gil de Paz et al$.$ 2007b) using ionized
gas shows an oxygen abundance of $\sim$Z$_{\odot}$/10 and (in the case
of M~83) a N/O abundance ratio close to Solar. This oxygen abundance
is compatible with a continuously-forming 1\,Gyr-old populations
(Boissier \& Prantzos 2000). The high N/O ratio found in M~83 can be
explained by either the recent infall of significant amounts of
pristine gas in regions having relatively high oxygen abundances and
N/O ratios or by the existence of a {\it top-light} Initial Mass
Function (IMF) in these (low-density) regions. A top-light IMF would
also explain the relatively paucity of HII regions (as seen in
H$\alpha$) in these disks. Detailed studies of a larger sample of XUV
disks are needed in order to determine whether or not these properties
are common to the outer edges of disks.

\begin{figure}[ht!]
\plotone{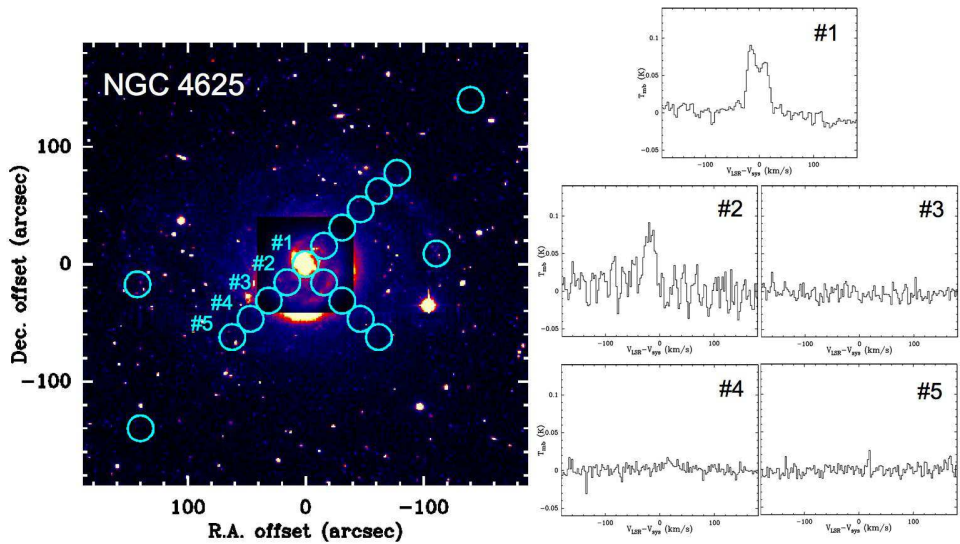}
\caption{IRAM 30-m $^{12}$CO(1-0) observations of the XUV-disk galaxy NGC~4625 (Gil de Paz et al$.$ 2005). {\it Left:} Deep optical ($R$-band) image of NGC~4625 with the positions observed. The size of the circles represents the size of the IRAM 30-m telescope main beam at 3\,mm. For sake of clarity the innermost region of the galaxy is shown with a different scaling. {\it Right:} Individual $^{12}$CO(1-0) spectra obtained along the galaxy semi-major axis.}
\end{figure}

\section*{Atomic and Molecular Gas Content}    
The outer regions of the XUV-disk galaxies observed to date show very
extended HI 21\,cm emission with its density peaks being associated
with the UV complexes seen by GALEX (see Figure~1). The amount of atomic
hydrogen in these reservoirs is large enough to maintain the current
level of star formation for several Gyrs. The possible contribution of
the molecular gas component is unknown since only the XUV disk
of NGC~4625 has been observed to date in CO to the required depth (see
Figure~2). Despite the tentative detection of CO at large
galactocentric distances in this object, the contribution of the
molecular component to the total gas seems to be low compared with that
from HI. Currently we cannot rule out the possibility that the HI associated with the
XUV disks is primarily due to photo-dissociation of H$_{2}$ by FUV photons
produced in the same stars that are responsible for the XUV emission.

\section*{CAHA-XUV Project: Exploring the Outer Edges of Spiral Galaxies}    
With the objective of determining whether or not the properties
described above, which are based almost exclusively on the detailed
analysis of the XUV disks of M~83 and NGC~4625, are common to the
entire population of galaxies showing XUV emission ($\sim$30\% of the
spiral galaxy population) we have recently started a large observing
program at the Calar Alto (CAHA) observatory (Almer\'{\i}a, Spain). We
will be obtaining deep optical ($UBRI$H$\alpha$) and near-infrared
($JK$) imaging of a sample of 65 nearby XUV disks. Follow-up
spectroscopy with CAHA 3.5-m and GTC 10.4-m is planned.

\acknowledgements 
We thank the Calar Alto Time Allocation Committee for the
generous allocation of time to the CAHA-XUV project.



\begin{thebibliography}{}
\bibitem[]{9}
Boissier, S., \& Prantzos, N. 2000, MNRAS, 312, 398
\bibitem[]{5}
Brook, C.B., et al. 2006, ApJ, 639, 126
\bibitem[]{12}
Ferguson, A., et al. 1998, ApJ, 506, L19
\bibitem[]{7}
Gil de Paz, A., et al. 2005, ApJ, 627, L29
\bibitem[]{1}
Gil de Paz, A., et al. 2007a, ApJS, 173, 185 
\bibitem[]{8}
Gil de Paz, A., et al. 2007b, ApJ, 661, 115
\bibitem[]{10}
Martin, D.C., et al. 2005, ApJ, 619, L1
\bibitem[]{3}
Martin, C., \& Kennicutt, R.C. 2001, ApJ, 555, 301
\bibitem[]{4}
Moll\'a, M., Ferrini, F., \& D\'{\i}az, A.I. 1996, ApJ, 466, 668
\bibitem[]{2}
Mu\~noz-Mateos, J.C., et al. 2007, ApJ, 658, 1006
\bibitem[]{6}
Thilker, D.A., et al. 2005, ApJ, 619, L79
\bibitem[]{11}
Thilker, D.A., et al. 2007, ApJS, 173, 538
\bibitem[]{13}
van Zee, L., et al. 1998, AJ, 116, 2805
\end{thebibliography}
\end{document}